\documentclass[11pt]{article}
\usepackage{latexsym}  
\usepackage{amsmath, amssymb} \usepackage{graphicx}
\usepackage{color}

\newcommand{\be}{\begin{equation}}  \newcommand{\ee}{\end{equation}}
\newcommand{\bea}{\begin{eqnarray}} \newcommand{\eea}{\end{eqnarray}}

\begin{document}
\begin{flushright}
YITP-15-36\\
May 11, 2015\\
Revised: July 7, 2015\\
Revised: September 29, 2015
\end{flushright}
\vspace{4\baselineskip}
\begin{center}
{\Large\bf Anapole moment of a chiral molecule revisited}
\end{center}
\vspace{1cm}
\begin{center}
{\large Takeshi Fukuyama$^{a,}$\footnote{E-mail:fukuyama@se.ritsumei.ac.jp}},
{\large Takamasa Momose$^{b,}$\footnote{E-mail:momose@chem.ubc.ca}}
and
{\large Daisuke Nomura$^{c,}$\footnote{E-mail:dnomura@yukawa.kyoto-u.ac.jp}}
\end{center}
\vspace{0.2cm}
\begin{center}
${}^{a}${\small \it Research Center for Nuclear Physics (RCNP),
             Osaka University, Ibaraki, Osaka, 567-0047, Japan}\\[.2cm]
${}^{b}${\small \it Department of Chemistry, 
  The University of British Columbia, Vancouver BC, V6T1Z1, Canada}\\[.2cm]
${}^{c}${\small \it Yukawa Institute for Theoretical Physics,
                    Kyoto University, Kyoto 606-8502, Japan}
\vskip 10mm
\end{center}
\vskip 10mm
\begin{abstract}
\noindent
Parity violation in a chiral, four-atom molecule is discussed. 
Given the geometrical positions of the four atoms, we calculate the
anapole moment of it. This problem was first discussed by Khriplovich
and Pospelov \cite{K-P}.
We give a detailed derivation for it so that it can be
more accessible to wider range of scientists. 
We correct errors in their results and generalize their initial state to 
$|s_{1/2}\rangle$ and $|p_{1/2}\rangle$ states.
We also discuss realistic candidates of the chiral molecules to which
this approach can be applied.
\end{abstract}
PACS numbers:  31.30.-i, 31.30.J-
  
\section{Introduction}

\noindent
Charge-conjugation (C), parity (P), and 
time-reversal (T) violations in atoms and molecules are very attractive
targets in searching for new physics beyond the standard model
(BSM physics) as well as for their own developments of atomic
and molecular physics and chemistry.
For molecules, we can utilize selectively the precise spectroscopies
and special environments of molecules suitable to determine
fundamental properties of particles. 
A typical example is the recent improvement of the upper bound
on the electron electric dipole moment (EDM) by using a ThO
molecule~\cite{ACME}. Also the measurements of energy difference
between left- and right-handed molecules are on-going~\cite{Quack_etal}.
Such situations conventionally enforce on us 
complicated numerical calculations 
based on {\it ab-initio} molecular orbital (MO) calculation
methods.
Since there are so many papers on {\it ab-initio} MO methods,
we cite only a textbook \cite{Quack} and 
a recent paper \cite{Thierfelder}. 
Though they are crucially important, it is rather difficult to elucidate the physical origin of symmetry violations in a given molecular system from molecular orbital calculations.
In order to remedy this deficit, analytical treatments of the same
objects as those studied by numerical calculations
have been awaited even with a cost of less rigor.
From this point of view, the work \cite{K-P} by Khriplovich and
Pospelov is methodologically very interesting.
Given the geometrical positions of four atoms, they calculated
the anapole moment of the chiral molecule which consists of
the four atoms.  There the 
P-violating (PV) anapole moment is directly and quantitatively related
with geometrical structure of molecules.

In this note we give a more detailed derivation of it
than in the original work~\cite{K-P},
so that it can be more accessible to many
physicists and chemists.
Their final result is also corrected.

We note here that in general, there are two possibilities concerning
the origin of parity violation in molecules.  One of the possibilities
is the weak interactions which intrinsically violate parity.  This
effect is known to work in nuclei in atomic systems, which is discussed,
for example, in Refs.~\cite{FK, FKS}.  The other possibility is parity
violation due to the geometric configuration of atoms in a molecule.  
When a molecule consists of more than four atoms, the molecule is not
necessarily superimposable on its mirror image.  Such a non-trivial
transformation property under parity is called the chirality, and the
chiral molecule and its mirror image are called enantiomers.  When we
take one of the enantiomers, parity is not a good symmetry to describe
the quantum state of the chiral molecule since parity is violated by
the configuration of the atoms.  It is this latter case of parity
violation which is discussed in Ref.~\cite{K-P} and which we consider in
this paper.

This paper is organized as follows. In Section 2, 
we describe our set-up of a four-atom molecule.  
In Section 3, we discuss the
perturbation due to the Coulomb interactions between
the valence electron and the light two atoms in the molecule.  In
Section 4, we calculate the anapole moment by using the
results obtained in Section 3.  We summarize
our results and give discussions in Section 5. 
In Appendix we discuss realistic candidates of the chiral
molecules to which this approach can be applied.
We work in the notation of the Landau-Lifshitz
textbooks~\cite{LandauLifshitz:QED, LandauLifshitz:QM}
unless noted otherwise.  We use the natural units in
which $\hbar=c=1$ throughout this paper.

\section{Model Set-up}

\begin{figure}[h]
\begin{center}
\includegraphics[scale=1.2]{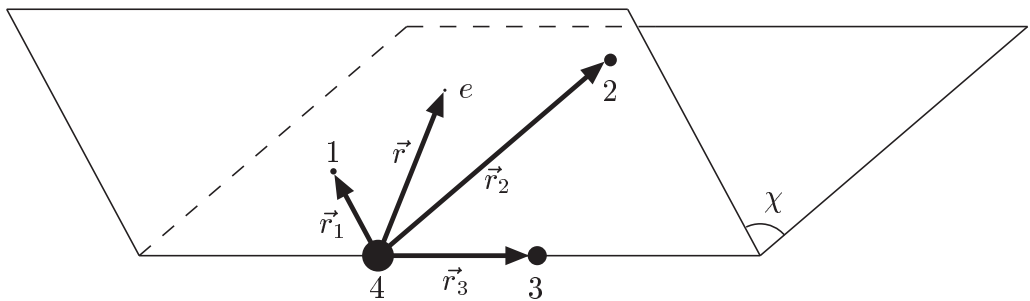}
\end{center}  
\vspace*{-0.5cm}
\caption{\label{fig:geom_config}
The configuration of the atoms in the four-atom
molecule we consider in this paper.  The numbers $j$ $(j=1, \ldots, 4)$
in the figure are the labels for the corresponding atoms $A_j$.
$e$ is a valence electron.  The two planes are spanned by
the atoms $A_1A_3A_4$ and the atoms $A_2A_3A_4$, respectively.
Also shown is the dihedral angle
$\chi$ of the molecule.}
\end{figure}

\noindent
Chirality in molecules first appears in four-atom molecules. 
Khriplovich and Pospelov considered a chiral molecule composed
of four atoms, $A_1,~A_2,~A_3,~A_4$, whose geometrical configuration
is given in Fig.~\ref{fig:geom_config} without specifying its
origin~\cite{K-P}.  
An assumption in this set-up is that $A_1$ and $A_2$ (whose
electric charges are $Z_1$ and $Z_2$ in units of the positron
charge $e$, respectively) are light
in comparison with $A_4$ and $A_3$. 
Throughout this paper, we take the position of $A_4$
as the origin of our coordinate system, and the direction
of $\vec{r}_3$ as the $z$-axis.

We first consider a valence electron which is captured
by the diatomic molecule $A_3A_4$.  
To specify the position of the valence electron, we
define the vector $\vec{r}$ whose initial and final points
are $A_4$ and the electron, respectively.
We assume that the electron is first in one of
the $p_{3/2}$ states with $\mu = \pm 3/2$, where $\mu$ is the
$z$ component of the total angular momentum.  We also assume that
the degeneracy in the energy levels of the atom $A_4$
is completely resolved by the presence of the atom $A_3$,
as assumed in Ref.~\cite{K-P}.
We treat the effect from the atoms $A_1$ and $A_2$ 
on the valence electron as a perturbation due to the
Coulomb potential, 
\be
V(\vec{r}) = -\frac{Z_1\alpha}{|\vec{r}-\vec{r}_1|}
       -\frac{Z_2\alpha}{|\vec{r}-\vec{r}_2|} ~.
\label{VPV}
\ee
Then, the Coulomb interactions between the valence electron
depicted as $e$ in Fig.~\ref{fig:geom_config} and the atoms
$A_1$ and $A_2$ can be treated as a perturbation to
the electron terms of the diatomic
molecule $A_3A_4$, which induces a PV anapole moment.
The unpaired electron is localized in the vicinity of $A_4$, and 
hence its orbital can be well described
in terms of the atomic orbitals of the atom $A_4$.
To lowest order of the approximation, the angular momentum
about the $A_3A_4$ axis is conserved.

\vspace*{0.2cm}

The final result we obtain in this paper is 
\begin{align}
\langle \vec{a} \rangle & =
\frac{\pi e}{m_e}\frac{4}{45}\left[r_1(s,p)-4r(s,p)\right]
 \frac{({\rm Ry})^2}{E_sE_p}Z_1Z_2\nonumber\\
& \quad \times 
(\vec{J} \cdot \vec{n}_3)
 (\vec{n}_1 \cdot [ \vec{n}_2\times \vec{n}_3 ] )
[  C_1(r_1) C_2(r_2) \vec{n}_2
 - C_1(r_2) C_2(r_1) \vec{n}_1] ~,
\label{final}
\end{align}
where $\vec{a}$ is the anapole moment operator
and $\langle \vec{a} \rangle$ its expectation value. 
${\rm Ry} \equiv \alpha^2 m_e /2$
is the Rydberg energy $(\simeq 13.6 {\text{eV}})$ 
with the electron mass $m_e$ and
$\vec{J}$ is the total angular momentum, 
and the definitions of $\vec{n}_{1,2,3}$, $C_{1,2}$, 
$r_1(s,p)$, $r(s,p)$, $E_{s,p}$ will be given in the next sections.
The result is different from that of Ref.~\cite{K-P} 
in the factor $4$ in front of $r(s,p)$ and also
in the overall factor by a factor of $2a_0$,
where $a_0$ is the Bohr radius, $a_0 \equiv 1/(m_e \alpha)$.  
Thus Eq.~(\ref{final}) explicitly shows the relation between
a PV observable and the geometrical structure of the molecule. 
In the subsequent sections we derive Eq.~(\ref{final}).

\section{Perturbation due to Coulomb Interactions}

\noindent
By the perturbation due to $V(\vec{r})$, the states
$|p_{3/2}, \mu=\pm 3/2 \rangle$ are slightly mixed
with the states $|s_{1/2}, \pm 1/2 \rangle$ 
and $|p_{1/2}, \pm 1/2 \rangle$.
First we write the potential $V(\vec{r})$ as
\begin{align}
 V(\vec{r}) &= 
   -Z_1\alpha
 \bigg\{ \theta(r-r_1) \left( 
    r_{1m} \frac{\partial}{\partial r_{m}} \frac1r
  + \frac12  r_{1m} r_{1n} \frac{\partial}{\partial r_{m}}
                           \frac{\partial}{\partial r_{n}} \frac1r
  + {\cal O}(r_1^3/r^3) \right) \nonumber \\
& \quad \quad \quad \quad + 
 \theta(r_1-r) \left( 
    r_{m} \frac{\partial}{\partial r_{1m}} \frac1{r_1}
  + \frac12  r_{m} r_{n} \frac{\partial}{\partial r_{1m}}
                         \frac{\partial}{\partial r_{1n}} \frac1{r_1}
  + {\cal O}(r^3/r_1^3) \right) \bigg\} \nonumber \\
& \quad
 - \bigg( (r_1, Z_1) \to (r_2, Z_2) \bigg)~, \label{eq:tmp3}
\end{align}
where $r_i \equiv |\vec{r}_i|$ $(i=1,2)$ and
the indices $n$ and $m$ $(n, m = 1, \ldots, 3)$ run over
the three components of the Cartesian coordinates, $(x, y, z)$. 
For later convenience, we integrate out the
dependence of the potential on $r ~ (\equiv |\vec{r}|)$,
and rewrite the potential Eq.~(\ref{eq:tmp3}) as
\begin{align}
 V(\vec{n}) &= {\rm Ry} \bigg( 2D_m n_m - 3Q_{mn} n_m n_n \bigg)~,
\label{eq:tmpFeb13_V}
\end{align}
where $\vec{n} \equiv \vec{r}/r$.  The quantities $D_m$
and $Q_{mn}$ are defined as
\begin{align}
 D_m    &\equiv Z_1 C_1(r_1) n_{1m} + Z_2 C_1(r_2)  n_{2m}~,
\label{eq:tmpFeb13_1}\\
 Q_{mn} &\equiv Z_1 C_2(r_1) n_{1m} n_{1n} 
             + Z_2 C_2(r_2) n_{2m} n_{2n} ~,
\label{eq:tmpFeb13_2}\\
 C_k(r_i) &\equiv a_0
  \int_0^\infty dr ~ r^2 R_{1/2}(r) R_{3/2}(r)
\bigg[ \frac{r^k_i}{r^{k+1}} \theta( r - r_i )
    +  \frac{r^k}{r_i^{k+1}} \theta( r_i - r ) \bigg] ~,
\label{eq:tmpFeb13}
\end{align} 
where $\vec{n}_i \equiv \vec{r}_i/r_i$ $(i=1,2)$. 
The factors $R_{3/2}(r)$ and $R_{1/2}(r)$ are the radial
parts of the $J=3/2$ and $J=1/2$ state wave functions for
the electron, respectively.
With these definitions, the quantity $C_k(r_i)$ becomes 
dimensionless.

As mentioned above, by the perturbation due to $V(\vec{r})$,
the states $|p_{3/2}, \mu=\pm 3/2 \rangle$ are slightly mixed
with other states $\psi_s$ and $\psi_p$ as:
\begin{align}
 |p_{3/2}, \mu=\pm 3/2 \rangle
 &\to  |p_{3/2}, \mu=\pm 3/2 \rangle
 + |\psi_s \rangle + |\psi_p \rangle ~.
 \label{eq:tmp14-1}
\end{align}
Here 
\begin{align}
 |\psi_s \rangle 
&= - \frac{2}{\sqrt{3}} \frac{{\rm Ry}}{E_s} D_i R_s(r)
  | s_{1/2}, \mu' \rangle ~,\label{eq:tmp38a} 
\end{align}
where $(i,\mu')=(+,1/2)$ and $(-,-1/2)$ for $\mu=3/2$
and $-3/2$, respectively, and
\begin{align}
 |\psi_p \rangle 
&= 
 \frac{2\sqrt{3}i}{5} \frac{{\rm Ry}}{E_p} Q_{mi} 
\sum_{\mu'=\pm 1/2} (\sigma_m )_{\alpha'\beta'}
R_p(r) | p_{1/2}, \mu' \rangle ~,
\label{eq:tmp38b}
\end{align}
where  $(i, \beta')=(+, 1), (-, 2)$ for $\mu=3/2$ and $-3/2$,
respectively, and 
$\alpha'=1, 2$ for $\mu'=1/2$ and $-1/2$, respectively.
The definitions of the ``$\pm$'' symbols which appear in the 
subscripts in Eqs.~(\ref{eq:tmp38a}) and (\ref{eq:tmp38b})
like $D_{\pm}$ will be given later in this paper at
Eq.~(\ref{eq:def_i_pm}).
The factors $E_s$ and $E_p$ are the energy levels of
the $|s_{1/2}, \mu'\rangle$ and $|p_{1/2}, \mu'\rangle$ states
$(\mu'=\pm 1/2)$,
measured from the state $|p_{3/2}, \mu \rangle$ $(\mu=\pm 3/2)$,
respectively, namely,
\begin{align}
%
 E_{s} \equiv E(s_{1/2},\mu') - E(p_{3/2}, \mu)~, ~~~~~~~~~
 E_{p} \equiv E(p_{1/2},\mu') - E(p_{3/2}, \mu)~.
\end{align}
In Eq.~(\ref{eq:tmp38b}), we have neglected the difference between
$E(p_{1/2},1/2)$ and $E(p_{1/2}, -1/2)$.
The functions $R_s(r)$ and $R_p(r)$ are the radial part
of the $s$- and $p$-wave-state wave functions, respectively.
The factor $\sigma_m$ $(m=1,\ldots,3)$ denotes the Pauli
matrices, and $(\sigma_m)_{\alpha'\beta'}$ $(\alpha',\beta'
=1, 2)$ its $(\alpha',\beta')$ component.
Note that our overall phase convention of $|p_{1/2}, \mu' \rangle$
differs from that of Ref.~\cite{K-P} by a factor
of $i$.  Since $\psi_s$ in Eq.~(\ref{eq:tmp14-1})
depends on the value of $\mu$, we denote
the $\psi_{s}$ which mixes with $|p_{3/2}, \mu= +3/2 \rangle$,
and $|p_{3/2}, \mu= -3/2 \rangle$
as $\psi_{s+}$ and $\psi_{s-}$, respectively. 
We also define $\psi_{p+}$ and $\psi_{p-}$ similarly, and
use the notation $\psi_{s\pm}$ and $\psi_{p\pm}$
instead of $\psi_s$ and $\psi_p$
where the distinction is necessary.
In the subsections just below, we prove Eqs.~(\ref{eq:tmp38a})
and (\ref{eq:tmp38b}).  

\subsection{Verification of Eq.~(\ref{eq:tmp38a})}

\noindent
In this subsection we verify Eq.~(\ref{eq:tmp38a}).
To do so, we need the solutions of 
the Dirac equation for an electron in the Coulomb potential,
$V(r)= - Z\alpha/r$.
Since we are interested only in the non-relativistic limit,
in this paper we consider only the upper
two components of the four-component Dirac spinor.

As a preparation, we first introduce some notations.
We define the two-component spinors $\chi_\alpha$ 
$(\alpha=\pm 1/2)$ as,
\begin{align}
 \chi_{1/2}  = \frac1{2\sqrt{\pi}}
             \begin{pmatrix} 1 \\ 0 \end{pmatrix}~,~~~~~~~~~
 \chi_{-1/2} = \frac1{2\sqrt{\pi}}
             \begin{pmatrix} 0 \\ 1 \end{pmatrix}~.
\label{eq:chi_def}
\end{align}
Namely, the angular parts of 
the states $|s_{1/2}, \alpha \rangle$ $(\alpha=\pm 1/2)$
are related to $\chi_\alpha$ by
\begin{align}
|s_{1/2}, \alpha \rangle = \chi_\alpha~,
\end{align}
where we have suppressed the radial part of the wave
function.  Below we suppress the radial part when there may
arise no confusion.
We also define the vectors $\vec{e}_i$ $(i=\pm, 0)$ as
\begin{align}
 \vec{e}_+ \equiv -i (1,  i, 0)/\sqrt{2}~,~~~~~~~~~~~~~
 \vec{e}_- \equiv  i (1, -i, 0)/\sqrt{2}~,~~~~~~~~~~~~~
 \vec{e}_0 \equiv  i (0, 0, 1)~.
\label{eq:eq_vec_e_pm}
\end{align}
They satisfy the relations,
\begin{align}
 \vec{e}_+ \cdot \vec{e}_+ = \vec{e}_- \cdot \vec{e}_- = 0 ~,~~~~~~~
 \vec{e}_+ \cdot \vec{e}_- = 1~, ~~~~~~~
 \vec{e}_{\pm} \cdot \vec{e}_0 = 0 ~,~~~~~~~
 \vec{e}_0 \cdot \vec{e}_0 = -1~. 
\end{align}
We denote the inner product of a general vector $\vec{k}$ and
$\vec{e}_{\pm, 0}$ as $k_{\pm, 0}$:
\begin{align}
 k_i \equiv  \vec{k} \cdot \vec{e}_i~, ~~~~~~~~~(i=\pm, 0).
\label{eq:def_i_pm}
\end{align}
With these definitions, the quantities $k_i$ $(i=\pm, 0)$
form a spherical tensor of rank 1.
Since the vectors $\vec{e}_i$
$(i=\pm, 0)$ are linearly independent of each other, 
any three dimensional vector $\vec{k}$ can be decomposed in
terms of $\vec{e}_i$ as
\begin{align}
 \vec{k} &=  (\vec{k} \cdot \vec{e}_-) \vec{e}_+
          + (\vec{k} \cdot \vec{e}_+) \vec{e}_-
          - (\vec{k} \cdot \vec{e}_0) \vec{e}_0 \nonumber\\
&=  k_- \vec{e}_+  + k_+ \vec{e}_-  - k_0 \vec{e}_0 ~.
\label{eq:tmp6}
\end{align}

Now, by using these notations, the angular part of the
$|p_{3/2}, \mu \rangle$ $(\mu = \pm 3/2)$ states 
can be written as
\begin{align}
|p_{3/2}, \mu \rangle = \sqrt{3} n_i \chi_\alpha~, 
\label{eq:P_32_n_chi}
\end{align}
where  $n_i = \vec{n} \cdot \vec{e}_i
 = (\vec{r}\cdot \vec{e_i})/r$ with $i=\pm$. 
The indices $i$ and $\alpha$ in the above equation should
be understood as $(i, \alpha )= (+, 1/2)$  for $\mu=3/2$, 
and $(i, \alpha)= (-, -1/2)$ for $\mu= -3/2$.

We are now ready to evaluate the angular integral
in the factor
$\langle s_{1/2}, 1/2 | 2 D_m n_m | p_{3/2}, 3/2 \rangle$,
where the integral is to be performed only over the 
angular variables.  The integral reads:
\begin{align}
\langle s_{1/2}, 1/2 | 2 D_m n_m | p_{3/2}, 3/2 \rangle
&=
 2 Z_1 C_1(r_1) \int d\Omega
 \frac1{2\sqrt{\pi}} 
 \frac{r_{1m}}{r_1} \frac{r_m}{r}
 (-i) \sqrt{\frac3{8\pi}} \sin\theta e^{i\phi} \nonumber\\
& \quad + \bigg((r_1,Z_1) \to (r_2,Z_2)\bigg) ~, 
\end{align}
where $d\Omega \equiv \sin\theta d\theta d \phi$.
The integral is straightforward, and the result is
\begin{align}
\langle s_{1/2}, 1/2 | 2 D_m n_m | p_{3/2}, 3/2 \rangle
&=
  2 Z_1 C_1(r_1) \frac{(-i)(n_{1x}+ in_{1y})}{\sqrt{6}}
+ \bigg((r_1,Z_1) \to (r_2,Z_2)\bigg)  \nonumber\\
&=
 \frac{2}{\sqrt{3}} \vec{D} \cdot \vec{e}_+
= \frac{2}{\sqrt{3}} D_+ ~.
\label{eq:tmp12}
\end{align}

Once the above matrix element is determined, the other matrix elements
$\langle s_{1/2}, \mu' | 2 D_m n_m | p_{3/2}, \mu \rangle$
for $(\mu,\mu')=(-3/2, \pm1/2), (3/2,-1/2)$ can also 
be determined by using the Wigner-Eckart theorem.  
To do so, first by using Eq.~(\ref{eq:tmp6}), we write
the quantity $D_m n_m$ as
\begin{align}
 D_m n_m &= D_- (\vec{e}_+ \cdot \vec{n}) 
          + D_+ (\vec{e}_- \cdot \vec{n}) 
          - D_0 (\vec{e}_0 \cdot \vec{n})  \nonumber \\ 
&= D_- n_+ + D_+ n_- - D_0 n_0 ~ .
\end{align}
Then we write the matrix elements as
\begin{align}
&\langle s_{1/2}, \mu' | 2 D_m n_m | p_{3/2}, \mu \rangle \nonumber\\
&= 2 D_- \langle s_{1/2}, \mu'| n_+ | p_{3/2}, \mu \rangle
+  2 D_+ \langle s_{1/2}, \mu'| n_- | p_{3/2}, \mu \rangle
-  2 D_0 \langle s_{1/2}, \mu'| n_0 | p_{3/2}, \mu \rangle
 \nonumber\\
&= 2 i (-1)^{3/2-\mu'} \langle s_{1/2}|| n || p_{3/2} \rangle 
\left\{
   D_- 
\begin{pmatrix}
  1/2  & 1 & 3/2 \\
 -\mu' & 1 & \mu
\end{pmatrix}
+  D_+ 
\begin{pmatrix}
  1/2  & 1  & 3/2 \\
 -\mu' & -1 & \mu
\end{pmatrix}
-  D_0 
\begin{pmatrix}
  1/2  & 1  & 3/2 \\
 -\mu' & 0  & \mu
\end{pmatrix}
\right\} \label{eq:Wigner-Eckart_apply}
\end{align}
where $\begin{pmatrix}
  j_1 & j_2 & j_3 \\
  m_1 & m_2 & m_3
\end{pmatrix}$ are the Wigner $3j$ symbols.  In the second
equality we have used the Wigner-Eckart theorem, 
\begin{align}
 \langle j' m'| f_{kq} |jm \rangle
= i^k (-1)^{j_{\text{max}}-m'} 
\begin{pmatrix}
  j' & k & j \\
 -m' & q & m
\end{pmatrix}
 \langle j' || f_{k} || j \rangle~,
\end{align}
where $j_{\text{max}} = \text{max}(j, j')$
and $f_{kq}$ $(q=-k, -k+1, \cdots, k)$ are a spherical
tensor of rank $k$. 
Since we already know the value of 
$\langle s_{1/2}, 1/2 | 2 D_m n_m | p_{3/2}, 3/2 \rangle$
from Eq.~(\ref{eq:tmp12}), 
we can determine the value of the double-line matrix element
$\langle s_{1/2}|| n || p_{3/2} \rangle$ by applying
Eq.~(\ref{eq:Wigner-Eckart_apply}) to the case $(\mu,\mu')=(3/2,1/2)$.
This can be done with the help of the numerical values of the 
Wigner $3j$ symbols as
\begin{align}
 \langle s_{1/2}|| n || p_{3/2} \rangle = i \frac{2}{\sqrt{3}}~.
\end{align}
Then, by using the value of $\langle s_{1/2}|| n || p_{3/2} \rangle$, 
the other relevant matrix elements can be determined from
Eq.~(\ref{eq:Wigner-Eckart_apply}) as
\begin{align}
\langle s_{1/2}, -1/2 | 2 D_m n_m | p_{3/2}, -3/2 \rangle
& =  \frac{2}{\sqrt{3}} D_- ~, \\
   \langle s_{1/2}, - 1/2 | 2 D_m n_m | p_{3/2}, + 3/2 \rangle
& = \langle s_{1/2}, + 1/2 | 2 D_m n_m | p_{3/2}, - 3/2 \rangle
  = 0~.
\end{align}

By combining all the above results, we arrive at
\begin{align}
\sum_{\mu'=\pm 1/2}
 | s_{1/2}, \mu' \rangle
\langle s_{1/2}, \mu' | 2 D_m n_m | p_{3/2}, \mu \rangle 
= \frac2{\sqrt{3}} D_i \chi_\alpha~,
\end{align}
where $i$ and $\alpha$ should be understood 
as $(i, \alpha )= (+, 1/2)$  for $\mu=3/2$, 
and $(i, \alpha)= (-, -1/2)$ for $\mu= -3/2$.
From this equation, Eq.~(\ref{eq:tmp38a}) immediately
follows.

\subsection{Verification of Eq.~(\ref{eq:tmp38b})}

\noindent
In this subsection we verify Eq.~(\ref{eq:tmp38b}).
The matrix element we would like to evaluate is,
\begin{align}
\langle p_{1/2}, \mu' | (-3)Q_{mn} n_m n_n | p_{3/2}, \pm 3/2 \rangle 
&=
- 3 Z_1 C_2(r_1) 
\langle p_{1/2}, \mu' | (\vec{n}_{1} \cdot \vec{n})^2
| p_{3/2}, \pm 3/2 \rangle \nonumber\\
& \quad + \bigg((r_1,Z_1) \to (r_2,Z_2)\bigg) ~.
\label{eq:tmp20}
\end{align}
Note that in our notation,
which is actually the notation of Landau-Lifshitz,
the states $|p_{1/2}, \mu \rangle$ $(\mu=\pm 1/2)$ are related to 
$\chi_\alpha$ $(\alpha=\pm 1/2)$ by
\begin{align}
 |p_{1/2}, \mu \rangle = -i ( \vec{\sigma} \cdot \vec{n}) \chi_\alpha~,
\end{align}
where $\alpha = 1/2$ and $-1/2$ for $\mu=1/2$ and $-1/2$, 
respectively.  The overall phase convention of $|p_{1/2}, \mu \rangle$
differs from that of Ref.~\cite{K-P} by a factor of $i$.

We first evaluate the factor
$\langle p_{1/2}, \mu' | (\vec{n}_1 \cdot \vec{n})^2 
| p_{3/2}, \pm 3/2 \rangle$,
where the integral is to be performed only over the 
angular variables.

To do so, we write the inner
product $\vec{n}_1 \cdot \vec{n}$ as
\begin{align}
\vec{n}_1 \cdot \vec{n}
&= 
  n_{1x} \frac{x}{r} + n_{1y} \frac{y}{r}
+ n_{1z} \frac{z}{r} \nonumber\\
&=
    n_{1x} \sin\theta\cos \phi 
  + n_{1y} \sin\theta\sin \phi 
  + n_{1z} \cos\theta        \nonumber\\
&=
\frac12 \sin\theta \bigg(
  (n_{1x}+in_{1y}) e^{-i\phi}
+ (n_{1x}-in_{1y}) e^{i\phi}
\bigg)   + n_{1z} \cos\theta ~. \label{eq:tmp15}
\end{align}
We now substitute the above expression into 
the matrix element 
$\langle p_{1/2}, 1/2 | (\vec{n}_1 \cdot \vec{n})^2 | p_{3/2}, 3/2 \rangle$
\begin{align}
\langle p_{1/2}, 1/2 | (\vec{n}_1 \cdot \vec{n})^2 | p_{3/2}, 3/2
 \rangle
&= 
 \int d\Omega
 \frac{i}{2\sqrt{\pi}} \cos\theta ~
( \vec{n}_1 \cdot \vec{n} )^2 ~
 (-i) \sqrt{\frac3{8\pi}} \sin\theta e^{i\phi} \nonumber\\
&=
  \int d\Omega
 \frac1{2\sqrt{\pi}} \cos\theta 
\bigg\{
\frac12 \sin\theta \bigg(
  (n_{1x}+in_{1y}) e^{-i\phi}
+ (n_{1x}-in_{1y}) e^{i\phi}
\bigg)   + n_{1z} \cos\theta 
\bigg\}^2 \nonumber\\
& \quad \quad \times
 \sqrt{\frac3{8\pi}} \sin\theta e^{i\phi} ~.
\end{align}
The angular integral is straightforward, and 
we are left with:
\begin{align}
\langle p_{1/2}, 1/2 | (\vec{n}_1 \cdot \vec{n})^2 | p_{3/2}, 3/2
 \rangle
&=
 \frac{\sqrt{2}}{5\sqrt{3}} 
\big( n_{1x}+in_{1y} \big) n_{1z} \nonumber \\
& = 
 \frac{2}{5\sqrt{3}} n_{1+} n_{10}~. \label{eq:tmp1}
\end{align}

We now determine the other relevant matrix elements
$\langle p_{1/2}, \pm 1/2 | (\vec{n}_1 \cdot \vec{n})^2 | p_{3/2}, \pm 3/2
\rangle$ by using the Wigner-Eckart theorem.  To do so,
we have to rewrite $(\vec{n}_1 \cdot \vec{n})^2$ in terms of
spherical tensors.
We construct a rank-2 spherical tensor $N_{2,q}$ 
$(q=-2,-1, \cdots, 2)$ by
combining two rank-1 spherical tensors $n_i$ $(i=\pm, 0)$.
This can be done by using the Clebsch-Gordan coefficients,
or equivalently the $3j$ symbols (see Eq.~(107.3) of 
Ref.~\cite{LandauLifshitz:QM}).  The results are:
\begin{align}
 N_{2,\pm2} = n^2_{\pm}~, ~~~~~~~~~~
 N_{2,\pm1} = \sqrt{2} n_{\pm} n_0~, ~~~~~~~~~~
 N_{2,0} =  \frac{2}{\sqrt{6}} n_+ n_-
 +  \frac{2}{\sqrt{6}} n_0^2~.
\end{align}
We also use a spherical tensor of rank 0, $N_{0,0}$,
which we construct
according to Eq.~(107.4) of Ref.~\cite{LandauLifshitz:QM},
\begin{align}
 N_{0,0} = 2 n_+ n_- - n_0^2~.
\end{align}
We now write $(\vec{n}_1 \cdot \vec{n})^2$ in terms of
$N_{2,q}$ and $N_{0,0}$:
\begin{align}
(\vec{n}_1 \cdot \vec{n})^2 &=
 \left( n_{1+} n_- +  n_{1-} n_+ - n_{10}  n_0\right)^2 \nonumber\\
&=
  (n_{1+})^2 N_{2,-2}
-\sqrt{2} n_{1+} n_{10} N_{2,-1}
+ c_{2,0} N_{2,0} 
-\sqrt{2} n_{1-} n_{10} N_{2,1}
+ (n_{1-})^2 N_{2,2}
+ c_{0,0} N_{0,0}~,
\end{align}
where $c_{2,0}$ and $c_{0,0}$ are linear combinations of
$n_{1+}n_{1-}$ and $n_{10} n_{10}$, whose explicit forms
we do not need for our purposes here.

We now apply the Wigner-Eckart theorem.  First we expand
$(\vec{n}_1 \cdot \vec{n})^2$ in terms of the tensors $N$,
\begin{align}
\langle p_{1/2}, \mu' | (\vec{n}_1 \cdot \vec{n})^2 | p_{3/2}, \mu
 \rangle
&= 
(n_{1+})^2 \langle p_{1/2}, \mu' | N_{2,-2} | p_{3/2}, \mu \rangle
- \sqrt{2} n_{1+}n_{10} 
\langle p_{1/2}, \mu' | N_{2,-1} | p_{3/2}, \mu \rangle
\nonumber\\
& \quad 
+ c_{2,0} \langle p_{1/2}, \mu' | N_{2,0} | p_{3/2}, \mu \rangle
- \sqrt{2} n_{1-}n_{10} 
\langle p_{1/2}, \mu' | N_{2,1} | p_{3/2}, \mu \rangle
\nonumber\\
& \quad 
+(n_{1-})^2 \langle p_{1/2}, \mu' | N_{2,2} | p_{3/2}, \mu \rangle
+ c_{0,0} \langle p_{1/2}, \mu' | N_{0,0} | p_{3/2}, \mu \rangle ~.
\end{align}
We use the Wigner-Eckart theorem to rewrite this as
\begin{align}
\langle p_{1/2}, \mu' | (\vec{n}_1 \cdot \vec{n})^2 | p_{3/2}, \mu
 \rangle
&= 
i^2 (-1)^{3/2-\mu'} 
\bigg\{
 (n_{1+})^2 \begin{pmatrix} 1/2   & 2 & 3/2\\
                           -\mu' & -2 & \mu \end{pmatrix}
- \sqrt{2} n_{1+} n_{10}
 \begin{pmatrix} 1/2  &  2 & 3/2\\
                -\mu' & -1 & \mu \end{pmatrix} \nonumber\\
& \quad \quad \quad \quad
+ c_{2,0}
 \begin{pmatrix} 1/2  & 2 & 3/2\\
                -\mu' & 0 & \mu \end{pmatrix}
- \sqrt{2} n_{1-} n_{10}
 \begin{pmatrix} 1/2  & 2 & 3/2\\
                -\mu' & 1 & \mu \end{pmatrix} \nonumber\\
& \quad \quad \quad \quad
+ (n_{1-})^2 \begin{pmatrix} 1/2 & 2 & 3/2\\
                           -\mu' & 2 & \mu \end{pmatrix} \bigg\}
\langle p_{1/2} || N_2 || p_{3/2} \rangle \nonumber\\
& \quad
+ (-1)^{3/2-\mu'} c_{0,0}
 \begin{pmatrix} 1/2  & 0 & 3/2\\
                -\mu' & 0 & \mu \end{pmatrix} 
\langle p_{1/2} || N_0 || p_{3/2} \rangle ~.
\end{align}
We can fix the value of $\langle p_{1/2} || N_2 || p_{3/2} \rangle$
from Eq.~(\ref{eq:tmp1}) as
\begin{align}
\langle p_{1/2} || N_2 || p_{3/2} \rangle = 
- \frac{2\sqrt{10}}{5\sqrt{3}}~.
\end{align}
By using this value, we can determine the following matrix
elements:
\begin{align}
\langle p_{1/2}, -1/2 | (\vec{n}_1 \cdot \vec{n})^2 
   | p_{3/2}, 3/2 \rangle 
&= i^2 (-1)^{3/2+1/2} (n_{1+})^2
 \begin{pmatrix} 1/2 & 2  & 3/2\\
                 1/2 & -2 & 3/2 \end{pmatrix}
\langle p_{1/2} || N_2 || p_{3/2} \rangle  \nonumber\\
&= 
 - \frac{2\sqrt{2}}{5\sqrt{3}} (n_{1+})^2 ~,  \label{eq:tmp31}\\
\langle p_{1/2}, 1/2 | (\vec{n}_1 \cdot \vec{n})^2 
   | p_{3/2}, -3/2 \rangle 
&= i^2 (-1)^{3/2-1/2} (n_{1-})^2 
 \begin{pmatrix} 1/2 & 2 & 3/2\\
                -1/2 & 2 & -3/2 \end{pmatrix}
\langle p_{1/2} || N_2 || p_{3/2} \rangle  \nonumber\\ 
&= 
  \frac{2\sqrt{2}}{5\sqrt{3}} (n_{1-})^2 ~,  \label{eq:tmp32}\\
\langle p_{1/2}, -1/2 | (\vec{n}_1 \cdot \vec{n})^2 
   | p_{3/2}, -3/2 \rangle 
&= i^2 (-1)^{3/2+1/2} (-\sqrt{2} n_{1-}n_{10})
 \begin{pmatrix} 1/2 & 2 &  3/2\\
                 1/2 & 1 & -3/2 \end{pmatrix}
\langle p_{1/2} || N_2 || p_{3/2} \rangle  \nonumber\\ 
&= - \frac{2}{5\sqrt{3}} n_{1-} n_{10} ~. \label{eq:tmp33}
\end{align}

The results Eqs.~(\ref{eq:tmp1}), (\ref{eq:tmp31}), (\ref{eq:tmp32}), 
(\ref{eq:tmp33}) can be compactly summarized as
\begin{align}
\langle p_{1/2}, \mu' | (\vec{n}_1 \cdot \vec{n})^2 | p_{3/2}, \mu
 \rangle
&=
\frac{2i}{5\sqrt{3}} n_{1i} (\vec{\sigma} 
\cdot \vec{n}_1)_{\alpha'\beta'}~,
\end{align}
where $(i, \beta')=(+, 1), (-, 2)$ for $\mu=3/2$ and $-3/2$,
respectively, and 
$\alpha'=1, 2$ for $\mu'=1/2$ and $-1/2$, respectively.
By substituting the above equation into Eq.~(\ref{eq:tmp20}),
we obtain
\begin{align}
\langle p_{1/2}, \mu' | (-3)Q_{mn} n_m n_n | p_{3/2}, \pm 3/2 \rangle 
=  
- \frac{2\sqrt{3}i}{5} Q_{im} (\sigma_m )_{\alpha'\beta'}~,
\end{align}
where $(i, \beta')=(+, 1), (-, 2)$ for $\mu=3/2$ and $-3/2$,
respectively, and 
$\alpha'=1, 2$ for $\mu'=1/2$ and $-1/2$, respectively.
It follows that, 
\begin{align}
\sum_{\mu'=\pm 1/2} 
| p_{1/2}, \mu'  \rangle
\langle p_{1/2}, \mu' | (-3)Q_{mn} n_m n_n | p_{3/2}, \pm 3/2 \rangle 
=  
- \frac{2\sqrt{3}}{5} Q_{im}  
\sum_{\mu'=\pm 1/2} (\sigma_m)_{\alpha'\beta'}
(\vec{\sigma} \cdot \vec{n}) \chi_{\mu'}~,
\end{align}
where $(i, \beta')=(+, 1), (-, 2)$ 
for $\mu=3/2$ and $-3/2$, respectively, and 
$\alpha'=1, 2$ for $\mu'=1/2$ and $-1/2$, respectively.

\section{Evaluation of Anapole Moment}

\noindent
We are now ready to calculate the anapole moment of the
four-atom molecule of Fig.~\ref{fig:geom_config}.

What we are interested in is the expectation value of
$\vec{a}$ sandwiched by the state represented by the
right-hand side of Eq.~(\ref{eq:tmp14-1}).
The non-trivial lowest-order contribution
comes from the terms,
\begin{align}
 \langle \vec{a} \rangle =
 \langle \psi_{s+} | \vec{a} | \psi_{p+} \rangle + {\text{(c.~c.)}}
 \label{eq:tmp2+}
\end{align}
%
for $\mu=3/2$, and
\begin{align}
 \langle \vec{a}  \rangle =
 \langle \psi_{s-} | \vec{a} | \psi_{p-} \rangle + {\text{(c.~c.)}}~,
  \label{eq:tmp2-}
\end{align}
for $\mu=-3/2$,
where $\text{(c.~c.)}$ stands for the complex conjugate.
We need a proof for the above statement (i.e.\ the
statement that the first-order perturbation and the rest of the
second-order perturbation both vanish), but at this moment,
we admit Eqs.~(\ref{eq:tmp2+}) and (\ref{eq:tmp2-}), and
calculate the right-hand sides of these equations.

The contribution $\langle \vec{a}_{\text{spin}} \rangle$
from the ``spin current'' to $\langle \vec{a} \rangle$ is,
\begin{align}
\langle \vec{a}_{\text{spin}} \rangle
&= - \frac{e\pi}{2m_e} \int d^3 \vec{r} ~ r^2
\left( \vec{\nabla} \times ( \psi_s^\dagger \vec{\sigma} \psi_p)
 \right) + \text{(c.~c.)}~.
\end{align}
We represent the integrand in terms of the spherical coordinates,
$(r,\theta,\phi)$:
\begin{align}
\langle \vec{a}_{\text{spin}} \rangle
 &=
 - \frac{e\pi}{2m_e} \int r^4 dr d\Omega 
\bigg\{
 \textbf{e}_r 
 \left( 
  \frac1{r} \partial_\theta  ( \psi_s^\dagger \sigma_\phi \psi_p)
+ \frac{\cos\theta}{r\sin\theta} 
   ( \psi_s^\dagger \sigma_\phi \psi_p)
- \frac{1}{r\sin\theta} \partial_\phi
   ( \psi_s^\dagger \sigma_\theta \psi_p)
 \right) \nonumber\\
& \quad \quad \quad \quad
+ \textbf{e}_\theta
 \left( 
  - \partial_r ( \psi_s^\dagger \sigma_\phi \psi_p)
  - \frac1r ( \psi_s^\dagger \sigma_\phi \psi_p)
  + \frac{1}{r\sin\theta} 
 \partial_\phi
   ( \psi_s^\dagger \sigma_r \psi_p)
 \right) \nonumber\\
& \quad \quad \quad \quad
+ \textbf{e}_\phi
 \left( 
   \partial_r ( \psi_s^\dagger \sigma_\theta \psi_p)
  + \frac1r ( \psi_s^\dagger \sigma_\theta \psi_p)
  - \frac{1}{r}  \partial_\theta
   ( \psi_s^\dagger \sigma_r \psi_p)
 \right) 
 \bigg\} + \text{(c.~c.)}~,
 \label{eq:tmp_deriv14_tmp1}
\end{align}
where the vectors $\textbf{e}_r, \textbf{e}_\theta, \textbf{e}_\phi$
are defined as
\begin{align}
\begin{pmatrix}
 \textbf{e}_r  \\  \textbf{e}_\theta \\  \textbf{e}_\phi
\end{pmatrix}
\equiv
\begin{pmatrix}
 \frac{\partial \vec{r}}{\partial r} \big/
\left| \frac{\partial \vec{r}}{\partial r} \right|   \\  
 \frac{\partial \vec{r}}{\partial \theta} \big/
\left| \frac{\partial \vec{r}}{\partial \theta} \right| \\
 \frac{\partial \vec{r}}{\partial \phi} \big/
\left| \frac{\partial \vec{r}}{\partial \phi} \right| 
\end{pmatrix}
=
\begin{pmatrix}
 \sin\theta \cos\phi & \sin\theta \sin\phi &  \cos\theta \\
 \cos\theta \cos\phi & \cos\theta \sin\phi & -\sin\theta \\
 - \sin\phi          & \cos\phi & 0
\end{pmatrix}
\begin{pmatrix}
 \textbf{e}_x  \\  \textbf{e}_y \\  \textbf{e}_z
\end{pmatrix}~,
\end{align}
where $\textbf{e}_x,  \textbf{e}_y, \textbf{e}_z$
are the unit vectors in the $x, y, z$ directions in the Cartesian
coordinates, respectively.
The matrices
$\sigma_r, \sigma_\theta, \sigma_\phi$ are the $2\times 2$
matrices defined in such a way that the identity below holds:
\begin{align}
 \vec{\sigma} \equiv
     \sigma_x \textbf{e}_x  +   \sigma_y \textbf{e}_y
   + \sigma_z \textbf{e}_z
 = 
     \sigma_r \textbf{e}_r  +   \sigma_\theta \textbf{e}_\theta
   + \sigma_\phi \textbf{e}_\phi~,
\end{align}
where $\sigma_{x,y,z}$ are the Pauli matrices.
The explicit forms of $\sigma_{r,\theta, \phi}$ are,
\begin{align}
 \sigma_r =
 \begin{pmatrix}
   \cos\theta        & e^{-i\phi}\sin\theta\\
  e^{i\phi}\sin\theta &  -\cos\theta
 \end{pmatrix}~, ~~~~~~~
 \sigma_\theta =
 \begin{pmatrix}
  - \sin\theta        & e^{-i\phi}\cos\theta\\
  e^{i\phi}\cos\theta &  \sin\theta
 \end{pmatrix}~, ~~~~~~~
 \sigma_\phi =
 \begin{pmatrix}
  0          & -i e^{-i\phi}\\
 i e^{i\phi} & 0
 \end{pmatrix}~.
\end{align}

We can greatly simplify Eq.~(\ref{eq:tmp_deriv14_tmp1})
by integrating-by-parts those terms which have $\partial_r$,
$\partial_\theta$, or $\partial_\phi$. 
First, integrating-by-parts those terms with $\partial_r$
is equivalent to replacing the operator $\partial_r$ with
a factor of $(-4/r)$:
\begin{align}
\langle \vec{a}_{\text{spin}} \rangle
&=
 - \frac{e\pi}{2m_e} \int r^4 dr d\Omega 
\bigg\{
 \textbf{e}_r 
 \left( 
  \frac1{r} \partial_\theta  ( \psi_s^\dagger \sigma_\phi \psi_p)
+ \frac{\cos\theta}{r\sin\theta} 
   ( \psi_s^\dagger \sigma_\phi \psi_p)
- \frac{1}{r\sin\theta} \partial_\phi
   ( \psi_s^\dagger \sigma_\theta \psi_p)
 \right) \nonumber\\
& \quad \quad \quad \quad
+ \textbf{e}_\theta
 \left( 
   \frac3r ( \psi_s^\dagger \sigma_\phi \psi_p)
  + \frac{1}{r\sin\theta} 
 \partial_\phi
   ( \psi_s^\dagger \sigma_r \psi_p)
 \right)
+ \textbf{e}_\phi
 \left( 
  - \frac3r ( \psi_s^\dagger \sigma_\theta \psi_p)
  - \frac{1}{r}  \partial_\theta
   ( \psi_s^\dagger \sigma_r \psi_p)
 \right) 
\bigg\} + \text{(c.~c.)}~.
\end{align}
We now integrate-by-parts the terms with $\partial_\theta$.
By noting that $\partial_\theta \textbf{e}_r = \textbf{e}_\theta$
and $\partial_\theta \textbf{e}_\phi = 0$, we obtain:
\begin{align}
\langle \vec{a}_{\text{spin}} \rangle
&=
 - \frac{e\pi}{2m_e} \int r^4 dr d\Omega 
\bigg\{
 \textbf{e}_r 
 \left( 
- \frac{1}{r\sin\theta} \partial_\phi
   ( \psi_s^\dagger \sigma_\theta \psi_p)
 \right)
+ \textbf{e}_\theta
 \left( 
   \frac2r ( \psi_s^\dagger \sigma_\phi \psi_p)
  + \frac{1}{r\sin\theta} 
 \partial_\phi
   ( \psi_s^\dagger \sigma_r \psi_p)
 \right) \nonumber\\
& \quad \quad \quad \quad
+ \textbf{e}_\phi
 \left( 
  - \frac3r ( \psi_s^\dagger \sigma_\theta \psi_p)
  + \frac{\cos\theta}{r\sin\theta}
   ( \psi_s^\dagger \sigma_r \psi_p)
 \right) 
\bigg\} + \text{(c.~c.)}~.
\end{align}
Finally, by integration-by-parts the terms with $\partial_\phi$,
by noting that $\partial_\phi \textbf{e}_r = \sin\theta \textbf{e}_\phi$
and $\partial_\phi \textbf{e}_\theta = \cos\theta \textbf{e}_\phi$,
we are left with:
\begin{align}
\langle \vec{a}_{\text{spin}} \rangle
&=
 - \frac{e\pi}{2m_e} \int r^4 dr d\Omega 
\bigg\{
 \textbf{e}_\theta
 \left( 
   \frac2r ( \psi_s^\dagger \sigma_\phi \psi_p)
 \right) 
+ \textbf{e}_\phi
 \left( 
  - \frac2r ( \psi_s^\dagger \sigma_\theta \psi_p)
 \right) 
\bigg\} + \text{(c.~c.)}~.
\end{align}
The integral with respect to $r$ is now easy:
\begin{align}
\langle \vec{a}_{\text{spin}} \rangle
=
 - \frac{e\pi}{m_e} r(s,p) \int d\Omega 
\bigg\{
 \textbf{e}_\theta
    ( \psi_s^\dagger \sigma_\phi \psi_p)_{\text{angular}}
- \textbf{e}_\phi
   ( \psi_s^\dagger \sigma_\theta \psi_p)_{\text{angular}}
\bigg\} + \text{(c.~c.)} ~, \label{eq:boxed_a_spin}
\end{align}
where the subscripts ``angular'' means that only the angular
part should be considered, forgetting about the radial part.
The factor $r(s,p)$ is defined in the unnumbered equation
just above Eq.~(15) of Ref.~\cite{K-P},
\begin{align}
 r(s,p) \equiv \int_0^\infty dr ~ r^3 R_s(r) R_p(r)~,
\end{align}
where $R_s(r)$ and $R_p(r)$ are the radial parts of
$\psi_s$ and $\psi_p$, respectively.

The contribution $\langle \vec{a}_{\text{orb}} \rangle$
from the usual ``orbital current'' to $\langle \vec{a} \rangle$ is,
\begin{align}
\langle \vec{a}_{\text{orb}} \rangle
&= - \frac{e\pi}{2m_e} \int d^3 \vec{r} ~ r^2
 \bigg\{ \psi_s^\dagger (-i) \vec{\nabla} \psi_p
      + i (\vec{\nabla} \psi_s^\dagger) \psi_p
 \bigg\} + \text{(c.~c.)}~.
\end{align}
The integral can be written in the spherical coordinates as:
\begin{align}
\langle \vec{a}_{\text{orb}} \rangle
&= - \frac{e\pi}{2m_e} \int r^4 dr d\Omega ~
 \bigg\{ \psi_s^\dagger (-i)
 \left(
   \textbf{e}_r \partial_r \psi_p
 + \textbf{e}_\theta \frac1r \partial_\theta \psi_p
 + \textbf{e}_\phi \frac1{r\sin\theta}
 \partial_\phi \psi_p \right) \nonumber\\
 & \quad \quad \quad \quad \quad \quad
 + i
 \left(
   \textbf{e}_r \partial_r \psi_s^\dagger
 + \textbf{e}_\theta \frac1r \partial_\theta \psi_s^\dagger
 + \textbf{e}_\phi \frac1{r\sin\theta}
 \partial_\phi \psi_s^\dagger\right) \psi_p
 \bigg\} + \text{(c.~c.)}~.
\end{align}
We can use $\partial_\phi \psi_s =\partial_\theta \psi_s = 0$
to drop two terms in the second line.
Upon integrating-by-parts those terms with $\partial_\theta$ or
$\partial_\phi$, we find that such terms cancel
with each other.  We are then left with:
\begin{align}
\langle \vec{a}_{\text{orb}} \rangle
&= i \frac{e\pi}{2m_e} \int r^4 dr d\Omega ~ \textbf{e}_r ~
 \bigg\{ \psi_s^\dagger ( \partial_r \psi_p)
 - (\partial_r \psi_s^\dagger) \psi_p
 \bigg\} + \text{(c.~c.)}~.
\end{align}
The $r$-integral is now easy:
\begin{align}
\langle \vec{a}_{\text{orb}} \rangle
= i \frac{e\pi}{2m_e} r_1(s,p) \int d\Omega ~ \textbf{e}_r ~
 \big(\psi_s^\dagger \psi_p \big)_{\text{angular}} + \text{(c.~c.)} ~,
 \label{eq:boxed_a_orb}
\end{align}
where $r_1(s,p)$ is defined as 
\begin{align}
 r_1(s,p) \equiv \int_0^\infty dr ~ r^4
 \left( R_s(r) \frac{dR_p(r)}{dr}
 - R_p(r) \frac{dR_s(r)}{dr} \right)~.
\end{align}

Before going further, it is convenient to calculate the
inner products
$\langle s_{1/2}, \mu=\pm1/2| \textbf{e}_r |p_{1/2}, \mu'=\pm1/2 \rangle$,
and $\langle s_{1/2}, \mu=\pm1/2|
(\textbf{e}_\theta \sigma_{\phi} - \textbf{e}_\phi \sigma_{\theta})
|p_{1/2}, \mu'=\pm1/2 \rangle$,
where the integral is to be performed only over the angular variables,
for all the possible combinations of $\mu$ and $\mu'$.

$\langle s_{1/2}, \mu=\pm1/2|\textbf{e}_r|
p_{1/2}, \mu'=\pm1/2 \rangle$ are evaluated to be, by using
the explicit solutions for the Dirac equation for a particle
in the Coulomb potential,
\begin{alignat}{3}
\langle s_{1/2}, +1/2 | \textbf{e}_r | p_{1/2}, +1/2 \rangle 
  &= -i \int d\Omega \frac1{4\pi}
  \begin{pmatrix} 1 & 0 \end{pmatrix}  \textbf{e}_r
  \begin{pmatrix} \cos\theta \\ \sin\theta e^{i\phi} \end{pmatrix}&
  \nonumber\\
 &= -i \int d\Omega \frac1{4\pi} \cos\theta
 ( \sin\theta \cos\phi \textbf{e}_x
  + \sin\theta \sin\phi \textbf{e}_y  + \cos\theta \textbf{e}_z )&
  \nonumber\\
  &= - \frac{i}3 \textbf{e}_z~. \label{eq:14-15}
\end{alignat}
Similarly,
\begin{alignat}{3} 
\langle s_{1/2}, +1/2 | \textbf{e}_r | p_{1/2}, -1/2 \rangle
 &= \int d\Omega \frac{-i}{4\pi} \sin\theta e^{-i\phi} 
 ( \sin\theta \cos\phi \textbf{e}_x
  + \sin\theta \sin\phi \textbf{e}_y  + \cos\theta \textbf{e}_z )&
 &= - \frac{i}3 ( \textbf{e}_x - i \textbf{e}_y )~,\label{eq:14-16}\\
\langle s_{1/2}, -1/2 | \textbf{e}_r | p_{1/2}, +1/2 \rangle
 &= \int d\Omega \frac{-i}{4\pi} \sin\theta e^{i\phi} 
 ( \sin\theta \cos\phi \textbf{e}_x
  + \sin\theta \sin\phi \textbf{e}_y  + \cos\theta \textbf{e}_z )&
 &= - \frac{i}3 ( \textbf{e}_x + i \textbf{e}_y )~,\label{eq:14-17}\\
 \langle s_{1/2}, -1/2 | \textbf{e}_r | p_{1/2}, -1/2 \rangle
 &= \int d\Omega \frac{i}{4\pi} \cos\theta
 ( \sin\theta \cos\phi \textbf{e}_x
  + \sin\theta \sin\phi \textbf{e}_y  + \cos\theta \textbf{e}_z )&
 & = + \frac{i}3 \textbf{e}_z~. \label{eq:14-18}
\end{alignat}

Similarly, the other elements $\langle s_{1/2}, \mu=\pm1/2|
(\textbf{e}_\theta \sigma_\phi-\textbf{e}_\phi \sigma_\theta)|
p_{1/2}, \mu'=\pm1/2 \rangle$ can be computed to be: 
\begin{align}
\langle s_{1/2}, +1/2|
(\textbf{e}_\theta \sigma_\phi-\textbf{e}_\phi \sigma_\theta)|
p_{1/2}, +1/2 \rangle
 &= + \frac23  \textbf{e}_z~, \label{eq:14-19}\\
\langle s_{1/2}, +1/2|
(\textbf{e}_\theta \sigma_\phi-\textbf{e}_\phi \sigma_\theta)|
p_{1/2}, -1/2 \rangle
 &= + \frac23  (\textbf{e}_x - i\textbf{e}_y)~, \label{eq:14-20}\\
\langle s_{1/2}, -1/2|
(\textbf{e}_\theta \sigma_\phi-\textbf{e}_\phi \sigma_\theta)|
p_{1/2}, +1/2 \rangle
 &= + \frac23  (\textbf{e}_x + i\textbf{e}_y)~,\label{eq:14-21}\\
\langle s_{1/2}, -1/2|
(\textbf{e}_\theta \sigma_\phi-\textbf{e}_\phi \sigma_\theta)|
p_{1/2}, -1/2 \rangle
 &= - \frac23  \textbf{e}_z~. \label{eq:14-22}
\end{align}

From Eqs.~(\ref{eq:14-15}--\ref{eq:14-22}),
the sum of the two contributions $\langle \vec{a}_{\text{spin}} \rangle$
(Eq.~(\ref{eq:boxed_a_spin})) and
$\langle \vec{a}_{\text{orb}} \rangle$ (Eq.~(\ref{eq:boxed_a_orb}))
can be expressed in a compact form:
\begin{align}
\langle \vec{a} \rangle
&= \langle \vec{a}_{\text{spin}} \rangle
 + \langle \vec{a}_{\text{orb}}  \rangle \nonumber\\
&= i \frac{e\pi}{2m_e} \bigg( r_1(s,p) - 4 r(s,p)\bigg)
 \int d\Omega ~ \textbf{e}_r ~ 
 \big(\psi_s^\dagger \psi_p \big)_{\text{angular}}
  + \text{(c.~c.)} ~. \label{eq:tmp27_Eq14LL}
\end{align}

We now evaluate the matrix element.   First, the 
contribution from $|s_{1/2}, 1/2 \rangle$ and
$|p_{1/2}, 1/2 \rangle$ to $\langle \psi_{s+} |\vec{a}|
\psi_{p+} \rangle+(\text{c.~c.})$ is, 
\begin{align}
& \left\{ i \int d\Omega (\psi_{s+}^\dagger \psi_{p+})_{\text{angular}}
 \textbf{e}_r + \text{(c.~c.)} \right\}
 \bigg|_{|s,1/2\rangle, |p,1/2\rangle} \nonumber\\
& \quad \quad =
i \frac{-i}{3} \textbf{e}_z
 \left\{ \bigg( - \frac2{\sqrt{3}} 
 \frac{{\rm Ry}}{E_s} Z_1 C_1(r_1) \frac{i(n_{1x}- in_{1y})}{\sqrt2}
\bigg) + \bigg((r_1,Z_1) \to (r_2,Z_2)\bigg) \right\} \nonumber\\
& \quad \quad \quad \times
 \left\{ \bigg( \frac{2\sqrt{3}i}{5}
 \frac{{\rm Ry}}{E_p} Z_1 C_2(r_1) n_{1z} \frac{-i(n_{1x} + in_{1y})}{\sqrt2}
\bigg) + \bigg((r_1,Z_1) \to (r_2,Z_2)\bigg) \right\} 
 + {\text{(c.~c.)}} \nonumber\\
&\quad \quad = \textbf{e}_z
 \frac4{15} \frac{({\rm Ry})^2}{E_sE_p} Z_1 Z_2
 \left[ \vec{n}_3 \cdot \bigg( \vec{n}_1 \times \vec{n}_2 \bigg)\right]
 \bigg\{   C_1(r_1) C_2(r_2) n_{2z}
         - C_2(r_1) C_1(r_2) n_{1z} \bigg\}~, \label{eq:tmp28_Eq14LL}
\end{align}
where we have used $\vec{n}_3 = \textbf{e}_z$.
If we further use $\vec{J} \cdot \vec{n}_3 = 3/2$
which holds for the case $\mu=3/2$, we get
\begin{align}
\left\{ i \int d\Omega (\psi_{s+}^\dagger \psi_{p+})_{\text{angular}}
 \textbf{e}_r + \text{(c.~c.)} \right\}
 \bigg|_{|s,1/2\rangle, |p,1/2\rangle}
&= \textbf{e}_z (\vec{J} \cdot \vec{n}_3)
 \frac8{45} \frac{({\rm Ry})^2}{E_sE_p} Z_1 Z_2
 \left[ \vec{n}_3 \cdot \bigg( \vec{n}_1 \times \vec{n}_2 \bigg)\right]
 \nonumber\\
&  \quad \times \bigg\{   C_1(r_1) C_2(r_2) n_{2z}
                        - C_2(r_1) C_1(r_2) n_{1z} \bigg\}~.
\end{align}

The contribution from $|s_{1/2}, 1/2 \rangle$ and
$|p_{1/2}, -1/2 \rangle$ to $\langle \psi_{s+} |\vec{a}|
\psi_{p+} \rangle+(\text{c.~c.})$ is, 
\begin{align}
& \left\{ i \int d\Omega (\psi_{s+}^\dagger \psi_{p+})_{\text{angular}}
 \textbf{e}_r + \text{(c.~c.)} \right\}
 \bigg|_{|s,1/2\rangle, |p,-1/2\rangle} \nonumber\\
& \quad \quad =
i \frac{-i}{3} (\textbf{e}_x -i \textbf{e}_y)
 \left\{ \bigg( - \frac2{\sqrt{3}} 
 \frac{{\rm Ry}}{E_s} Z_1 C_1(r_1) \frac{i(n_{1x}- in_{1y})}{\sqrt2}
\bigg) + \bigg((r_1,Z_1) \to (r_2,Z_2)\bigg) \right\} \nonumber\\
& \quad \quad \quad \times
 \left\{ \bigg( \frac{2\sqrt{3}i}{5}
 \frac{{\rm Ry}}{E_p} Z_1 C_2(r_1) (n_{1x}+in_{1y})
 \frac{-i(n_{1x} + in_{1y})}{\sqrt2}
\bigg) + \bigg((r_1,Z_1) \to (r_2,Z_2)\bigg)  \right\} 
 + {\text{(c.~c.)}} \nonumber\\
& \quad \quad =
 \frac{4({\rm Ry})^2}{15E_sE_p} Z_1 Z_2
 \left[ \vec{n}_3 \cdot \bigg( \vec{n}_1 \times \vec{n}_2 \bigg)\right]
 \bigg\{   C_1(r_1) C_2(r_2) 
   (n_{2x} \textbf{e}_x +  n_{2y} \textbf{e}_y ) 
         - C_2(r_1) C_1(r_2) 
   (n_{1x} \textbf{e}_x +  n_{1y} \textbf{e}_y ) 
 \bigg\} \nonumber\\
& \quad \quad \quad
 + \frac{4({\rm Ry})^2}{15 E_sE_p}
 \bigg[ 
     Z_1^2  C_1(r_1) C_2(r_1) \big( \vec{n}_1 \times \vec{n}_3 \big)
     \big| \vec{n}_1 \times \vec{n}_3 \big|^2 \nonumber\\
& \qquad \qquad \qquad \qquad
  + Z_1 Z_2 C_1(r_1) C_2(r_2) \big( \vec{n}_2 \times \vec{n}_3 \big)
 \bigg( ( \vec{n}_1 \times \vec{n}_3 ) \cdot
        ( \vec{n}_2 \times \vec{n}_3 ) \bigg) \nonumber\\
& \qquad \qquad \qquad \qquad
  + Z_1 Z_2 C_1(r_2) C_2(r_1) \big( \vec{n}_1 \times \vec{n}_3 \big)
   \bigg( ( \vec{n}_1 \times \vec{n}_3 ) \cdot
        ( \vec{n}_2 \times \vec{n}_3 ) \bigg) \nonumber\\
& \qquad \qquad \qquad \qquad
  +  Z_2^2  C_1(r_2) C_2(r_2) \big( \vec{n}_2 \times \vec{n}_3 \big)
     \big| \vec{n}_2 \times \vec{n}_3 \big|^2
\bigg]~. \label{eq:tmp31_Eq14LL}
\end{align}

By substituting the sum of
Eqs.~(\ref{eq:tmp28_Eq14LL}) and (\ref{eq:tmp31_Eq14LL})
into Eq.~(\ref{eq:tmp27_Eq14LL}), we obtain, for $\mu=3/2$,
\begin{align}
\langle \vec{a} \rangle
 &= \frac{e\pi}{m_e} \bigg( r_1(s,p) - 4 r(s,p)\bigg)
  \frac{4 ({\rm Ry})^2}{45 E_sE_p} Z_1 Z_2 (\vec{J} \cdot \vec{n}_3) 
 \left[ \vec{n}_3 \cdot \bigg( \vec{n}_1 \times \vec{n}_2 \bigg)\right]
 \bigg\{   C_1(r_1) C_2(r_2) \vec{n}_{2}
         - C_2(r_1) C_1(r_2) \vec{n}_{1} \bigg\} \nonumber\\
& \quad + \frac{e\pi}{m_e} \bigg( r_1(s,p) - 4 r(s,p)\bigg)
 \frac{4({\rm Ry})^2}{45 E_sE_p} (\vec{J} \cdot \vec{n}_3)
 \bigg[ 
     Z_1^2  C_1(r_1) C_2(r_1) \big( \vec{n}_1 \times \vec{n}_3 \big)
     \big| \vec{n}_1 \times \vec{n}_3 \big|^2 \nonumber\\
& \qquad \qquad \qquad \qquad
  + Z_1 Z_2 C_1(r_1) C_2(r_2) \big( \vec{n}_2 \times \vec{n}_3 \big)
 \bigg( ( \vec{n}_1 \times \vec{n}_3 ) \cdot
        ( \vec{n}_2 \times \vec{n}_3 ) \bigg) \nonumber\\
& \qquad \qquad \qquad \qquad
  + Z_1 Z_2 C_1(r_2) C_2(r_1) \big( \vec{n}_1 \times \vec{n}_3 \big)
   \bigg( ( \vec{n}_1 \times \vec{n}_3 ) \cdot
        ( \vec{n}_2 \times \vec{n}_3 ) \bigg) \nonumber\\
& \qquad \qquad \qquad \qquad
  +  Z_2^2  C_1(r_2) C_2(r_2) \big( \vec{n}_2 \times \vec{n}_3 \big)
     \big| \vec{n}_2 \times \vec{n}_3 \big|^2
\bigg]~, \label{eq:tmp32_Eq14Ll}
\end{align}
where we have used $\vec{J} \cdot \vec{n}_3 = 3/2$
which holds for the case $\mu=3/2$.
The terms in the second and the subsequent lines are
all perpendicular to the axis of the diatomic molecule $A_3A_4$
and vanish by averaging the orientation of the molecule with this axis fixed.
As long as the rotational symmetry around the $z$-axis 
is a good symmetry,
this averaging is legitimate since there is a quantum-mechanical
uncertainty relation between $J_z$ and the azimuthal angle $\phi$
which describes the orientation of the
molecule with the $A_3A_4$ axis fixed at the $z$-axis.
Thus we obtain Eq.~(\ref{final}).   The same conclusion can also
be obtained for $\mu=-3/2$.

In the above discussion, we have neglected the 
possible contribution from the terms at first order 
of perturbation.  Below we show that this can be justified.
At first order of perturbation, the possible contribution
to $\langle \vec{a} \rangle$ comes from
\begin{align}
 \langle \vec{a} \rangle = \langle \psi_{s \pm} | \vec{a} 
   | p_{3/2}, \mu=\pm 3/2 \rangle + (\text{c.~c.})~
\label{eq:tmp1a} 
\end{align}
(The matrix element of $\vec{a}$
between $|\psi_{p \pm}\rangle$ and $|p_{3/2} \rangle$ vanishes
because of parity).  To evaluate Eq.~(\ref{eq:tmp1a}), we can use
\begin{align}
 \langle \vec{a} \rangle & = 
    \langle \vec{a}_{\text{spin}} \rangle
  + \langle \vec{a}_{\text{orb}} \rangle
\\
\langle  \vec{a}_{\text{spin}}  \rangle
&= - \frac{e\pi}{m_e} r(s,p) \int d\Omega 
\bigg\{
 \textbf{e}_\theta
    ( \psi_s^\dagger \sigma_\phi \psi_p)_{\text{angular}}
- \textbf{e}_\phi
   ( \psi_s^\dagger \sigma_\theta \psi_p)_{\text{angular}}
\bigg\} + \text{(c.~c.)} ~, \label{eq:boxed_a_spin_2} \\
\langle  \vec{a}_{\text{orb}} \rangle
&= i \frac{e\pi}{2m_e} r_1(s,p) \int d\Omega ~ \textbf{e}_r ~
 \big(\psi_s^\dagger \psi_p \big)_{\text{angular}} + \text{(c.~c.)} ~,
 \label{eq:boxed_a_orb_2}
\end{align}
where $\psi_p$ should be understood as $|p_{3/2}, \mu=\pm3/2\rangle$.
The integrals which can appear in Eqs.~(\ref{eq:boxed_a_spin_2}) and
(\ref{eq:boxed_a_orb_2}) are:
\begin{align}
 \langle s_{1/2}, +1/2 | \textbf{e}_r | p_{3/2}, +3/2 \rangle 
 &= \frac{-i}{\sqrt6} (\textbf{e}_x + i\textbf{e}_y)~,
\label{eq:tmp_int6} \\
\langle s_{1/2}, -1/2 | \textbf{e}_r | p_{3/2}, -3/2 \rangle 
 &=  \frac{i}{\sqrt6} (\textbf{e}_x - i\textbf{e}_y)~,
\label{eq:tmp_int7} \\
   \langle s_{1/2}, +1/2 | \textbf{e}_r | p_{3/2}, -3/2 \rangle
&= \langle s_{1/2}, -1/2 | \textbf{e}_r | p_{3/2}, +3/2 \rangle = 0~,
\label{eq:tmp_int8}
\end{align}
and
\begin{align}
 \langle s_{1/2}, +1/2 | 
  ( \textbf{e}_\theta \sigma_{\phi} - \textbf{e}_\phi \sigma_{\theta})
   | p_{3/2}, +3/2 \rangle 
 &= \frac{-1}{\sqrt6} (\textbf{e}_x + i\textbf{e}_y)~, 
\label{eq:tmp_int9}\\
 \langle s_{1/2}, -1/2 | 
  ( \textbf{e}_\theta \sigma_{\phi} - \textbf{e}_\phi \sigma_{\theta})
   | p_{3/2}, -3/2 \rangle 
 &= \frac{1}{\sqrt6} (\textbf{e}_x - i\textbf{e}_y)~, 
\label{eq:tmp_int10} \\
 \langle s_{1/2}, +1/2 | 
  ( \textbf{e}_\theta \sigma_{\phi} - \textbf{e}_\phi \sigma_{\theta})
   | p_{3/2}, -3/2 \rangle 
&=
 \langle s_{1/2}, -1/2 | 
  ( \textbf{e}_\theta \sigma_{\phi} - \textbf{e}_\phi \sigma_{\theta})
   | p_{3/2}, +3/2 \rangle  = 0 ~.
\label{eq:tmp_int11}
\end{align}

From Eqs.~(\ref{eq:tmp_int6})--(\ref{eq:tmp_int11}), we find that
the vector
$\langle \psi_{s \pm} | \vec{a} | p_{3/2}, \mu=\pm 3/2 \rangle
+ (\text{c.~c.})$ is zero or perpendicular to $\vec{n}_3$,
namely, the axis of the diatomic molecule $A_3A_4$.  In this case,
this vector vanishes when averaged over the orientation of
the molecule with $\vec{n}_3$ being fixed.  Therefore, to
first order of perturbation, there is no contribution to the
anapole.  The same comment also applies to the contribution
which comes from second order of perturbation like the term
\begin{align}
  \langle \vec{a} \rangle = \langle \psi'_{s \pm} | \vec{a} 
   | p_{3/2}, \mu=\pm 3/2 \rangle + (\text{c.~c.})~,
\end{align}
where $|\psi'_{s} \rangle$ is the $s$-wave state which appears
as a second-order correction to the initial wave function
$|p_{3/2}\rangle$.  (Another second-order contribution 
$\langle \psi'_{p} | \vec{a} 
| p_{3/2}, \mu=\pm 3/2 \rangle + (\text{c.~c.})$,
where $\psi'_{p}$ is the $p$-wave state which appears
as a second-order correction to the initial wave function
$|p_{3/2}\rangle$, vanishes from parity conservation.)
In this case, the angular integrals
involved are Eqs.~(\ref{eq:tmp_int6}), (\ref{eq:tmp_int7}), 
(\ref{eq:tmp_int8}), (\ref{eq:tmp_int9}), (\ref{eq:tmp_int10}), 
and (\ref{eq:tmp_int11}), which all vanish when averaged
over the orientation of the molecule with $\vec{n}_3$ being fixed.
Therefore there is no other second-order contribution 
to Eq.~(\ref{final}) which does not vanish after the
average over the orientations of the molecule with fixed
$\vec{n}_3$ as mentioned above.

\section{Discussion}

\noindent
We have analytically derived the direct relation between the
anapole moment and the geometrical structure of chiral molecules.
A four-atom molecule has been studied as an example since it is the 
simplest molecule having chirality.
We have focused on a valence electron which is captured
by the diatomic molecule $A_3A_4$, and treated effects
from the other atoms $A_1$ and $A_2$ on the electron
as perturbation.
We have computed the corrections to the wave function of the
electron in the $|p_{3/2}, \mu=\pm 3/2\rangle$ state
by the Coulomb interactions from the atoms $A_1$ and
$A_2$, and calculated the anapole moment.
Though this method was first developed in Ref.~\cite{K-P},
this method has been discussed in much more detail in this paper
to be accessible
to many scientists over wider regions and
their final result has been corrected.
The electron terms for $n$-atom molecules ($n\leq 3$) reserve the
symmetry of molecules.  This is not the case for $n\geq 4$ and
needs complicated processes~\cite{Quack, Thierfelder}.
On the other hand, the procedures developed in this paper
have given the direct relation between the PV interactions and
the geometrical structure
without wandering into complicated {\it ab-initio} MO calculations. 
Of course, these two approaches
are complementary and we need some bridge between the two approaches.
One point which may need improvement in our approach is that 
our approach might seem to depend on the peculiar initial state
$|p_{3/2}\rangle$.
We may generalize it to the cases where the initial state
is $|p_{1/2}\rangle$ or $|s_{1/2}\rangle$ state.
Starting from $|s_{1/2}\rangle$ state in place of $|p_{3/2}\rangle$,
we obtain
\begin{align}
 \langle \vec{a} \rangle 
&=
  \frac{2e\pi}{15m_e} \bigg( r_1(s,p) - 4r(s,p) \bigg)
\left(
  \frac1{E(s_{1/2})-E(p_{3/2}, 3/2)}
- \frac1{E(s_{1/2})-E(p_{3/2}, -1/2)}
\right) \frac{(\text{Ry})^2}{E(s_{1/2})-E(p_{1/2})} \nonumber\\
& \quad \quad 
\times
 Z_1 Z_2 
 \left[ \vec{n}_3 \cdot \bigg( \vec{n}_1 \times \vec{n}_2 \bigg)\right]
 \bigg\{   C_1(r_1) C_2(r_2) \vec{n}_2
         - C_2(r_1) C_1(r_2) \vec{n}_1 \bigg\}
\nonumber\\
& \quad 
 - \frac{4 e\pi}{15 m_e} 
  \bigg( \tilde{r}_1(s,p) + 2\tilde{r}(s,p) \bigg)
  \frac1{E(s_{1/2}, 1/2)-E(p_{1/2}, -1/2)}
  \frac{(\text{Ry})^2}{E(s_{1/2}, 1/2)-E(p_{3/2}, 1/2)} \nonumber\\
& \quad \quad 
\times
 Z_1 Z_2 
 \left[ \vec{n}_3 \cdot \bigg( \vec{n}_1 \times \vec{n}_2 \bigg)\right]
 \bigg\{   \tilde{C}_1(r_1) C_2(r_2) \vec{n}_2
         - C_2(r_1) \tilde{C}_1(r_2) \vec{n}_1 \bigg\}~,
\label{s1/2}
\end{align}
where $\tilde{C}_1(r_i) \equiv a_0 \int_0^\infty dr \ r^2 R_{s_{1/2}}(r)
R_{p_{1/2}}(r) [ (r_i/r^2)\theta(r-r_i)
+ (r/r_i^2)\theta(r_i-r)]$, where $R_{s_{1/2}}(r)$ and 
$R_{p_{1/2}}(r)$ are the radial wave functions of the initial
$s_{1/2}$ state and the intermediate $p_{1/2}$ state, respectively.
$\tilde{r}(s,p)$ and $\tilde{r}_1(s,p)$ are defined as
$\tilde{r}(s,p) \equiv \int_0^\infty dr
 r^3 R_{s_{1/2}}(r) R_{p_{3/2}}(r)$
and 
$\tilde{r}_1(s,p) \equiv \int_0^\infty dr
 r^4 (R_{s_{1/2}}(r) R'_{p_{3/2}}(r) - R'_{s_{1/2}}(r) R_{p_{3/2}}(r))
$, where $R_{p_{3/2}}(r)$ is 
the radial wave function of the $p_{3/2}$ state.
Instead, if we start from $|p_{1/2}\rangle$ state, we obtain
\begin{align}
 \langle \vec{a} \rangle 
&=
 + \frac{2e\pi}{15m_e} \bigg( r_1(s,p) - 4r(s,p) \bigg)
\left(
  \frac1{E(p_{1/2})-E(p_{3/2}, 3/2)}
- \frac1{E(p_{1/2})-E(p_{3/2}, -1/2)}
\right) \frac{(\text{Ry})^2}{E(p_{1/2})-E(s_{1/2})} \nonumber\\
& \quad \quad 
\times
 Z_1 Z_2 
 \left[ \vec{n}_3 \cdot \bigg( \vec{n}_1 \times \vec{n}_2 \bigg)\right]
 \bigg\{   C_1(r_1) C_2(r_2) \vec{n}_2
         - C_2(r_1) C_1(r_2) \vec{n}_1 \bigg\} \nonumber\\
& \quad
+  \frac{4 e\pi}{15 m_e} \bigg( \tilde{r}_1(s,p) + 2\tilde{r}(s,p) \bigg) 
  \frac1{E(p_{1/2}, 1/2) -E(s_{1/2}, -1/2)} 
  \frac{(\text{Ry})^2}{E(p_{1/2}, 1/2) -E(p_{3/2}, -1/2)} \nonumber\\
& \quad \times
Z_1 Z_2 
 \left[ \vec{n}_3 \cdot \bigg( \vec{n}_1 \times \vec{n}_2 \bigg)\right]
 \bigg\{   \tilde{C}_1(r_1) C_2(r_2) \vec{n}_2
         - C_2(r_1) \tilde{C}_1(r_2) \vec{n}_1 \bigg\} ~.
\label{eq:p1/2}
\end{align}
This is of a similar form to Eq.~(\ref{s1/2}).

The second point which needs discussions is to apply this
approach to realistic molecules.  This is discussed in Appendix.

This approach has many applications.
One of them is the evaluation of the energy differences
between optical isomers. 
The PV potential due to the weak neutral boson exchange
is given by \cite{zeldovich2, Sandars, Khriplovich:1980}
\begin{align}
V^{PV} = \frac{G_F}{4\sqrt{2} m_e}\sum_{\alpha,i} Q_W^\alpha
   \{ \boldsymbol{\sigma}_i \cdot {\bf p}_i, 
      \delta^{(3)}({\bf r}_{i\alpha}) \}_+ ~,
\label{S1}
\end{align}
where $Q_W^{\alpha}$ is the weak charge of the $\alpha$-th atom
in a molecule:
\begin{align}
Q_W^\alpha = (1-4\sin^2\theta_W)Z^\alpha-N^\alpha ~,
\end{align}
with the atomic number $Z^{\alpha}$ and the neutron number $N^{\alpha}$
of the $\alpha$-th atom, where $\alpha$ runs over the atoms composing
of the molecule and $i$ labels the electrons. The precedent analytical
methods have not discussed the direct relation between the energy
differences and geometrical structures.  The model set-up is the same as
in the present paper and we may replace the anapole moment by
Eq.~(\ref{S1}).
The detail will be discussed in a separate form~\cite{F-N2}.

\section*{Acknowledgements}

\noindent
The work of T.~F.\ is supported in part by the Grant-in-Aid for
Science Research from the Ministry of Education, Science and
Culture of Japan (No.\ 26247036) and by JSPS-INSA Bilateral
Joint Research Projects 2012-2016.
D.~N.\ is a Yukawa Fellow, and this work was partially
supported by the Yukawa Memorial Foundation.

\appendix

\section{Candidate molecules}
The present paper discusses that chiral molecules having an unpaired
electron would be good systems for the measurement of the anapole
moment, and the analytical derivation of the expectation value of the
anapole moment Hamiltonian (Eqs.~(47) and (57)) for four-atom molecules,
which are the smallest molecules that have chirality, is given.
In order to further simplify the analytical derivation, we assumed
that the nature of the unpaired electron is mainly described by atomic
orbitals attached to $A_4$ (Fig.1), and the projection of the angular
momentum along the $A_4-A_3$ axis is a good quantum number.  Molecules
that  satisfy these conditions may be those whose $A_4$ is a heavy atom,
in which most of the unpaired electron distributes on $A_4$, and whose
$A_1$ and $A_2$ are lighter atoms than $A_3$ and $A_4$.

An example of a four-atom molecule with an unpaired electron is the
HNOH radical, in which one hydrogen of  an amine group is removed
from hydroxylamine (NH$_2$OH).  However, it is not a chiral molecule
because this radical has a planer equilibrium structure \cite{Wu:2009be}. 
We have searched for various four-atom molecules similar to the HNOH
radical by the use of {\it ab-initio} molecular orbital calculations,
and found several molecules that have a chiral structure.
They include the FAsSH radical, whose dihedral angle, $\chi$, at the
equilibrium structure is 61.3 degrees.  The potential surface of this
molecule along the dihedral angle, $\chi$, based on 
MP2/6-31G(d) \cite{g09} is shown in Fig.2.  The potential barrier
along $\chi$ is 354.6 cm$^{-1}$, while the vibrational frequency
along $\chi$ under the harmonic approximation is 136.4 cm$^{-1}$.
The gross orbital population analysis of a simple ROHF molecular orbital
calculation indicates that about 82 \% of the electron spin locates on
the $4p$ orbital of the As atom.  Since the ground state of AsS is
$^2\Pi_{1/2}$, the anapole moment of this radical may be roughly
approximated by Eq.~(\ref{eq:p1/2}). 

Another example is the FPOH radical, which has the dihedral angle, 
$\chi$, of 53.0 degrees.  The electron spin of this molecule, however,
distributes widely over the radical, and therefore it will be necessary
to include higher order terms that we ignored in this paper to estimate
the anapole moment of this radical.

So far a spectroscopic study of the HNOH radical has been reported
by matrix isolation spectroscopy in solid hydrogen \cite{Wu:2009be}.
No spectroscopic information were reported for other molecules.

 \begin{figure}[t]
\begin{center}
\includegraphics[width=0.40\textwidth]{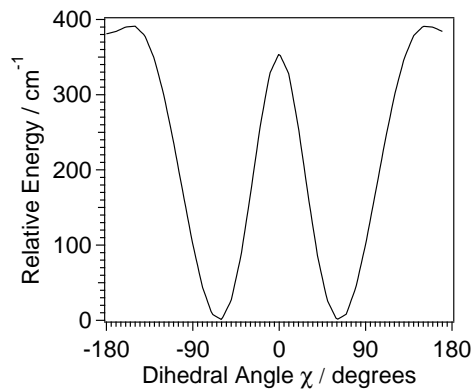}
\caption{\sl \small 
The potential curve of FAsSH along the F-As-S-H dihedral angle obtained by an MP2/6-31G(d) {\it ab-initio} molecular orbital calculation.}\label{fig2}
\end{center}
\end{figure}

\end{document}